\documentstyle[preprint,aps,epsfig]{revtex}

\preprint{\begin{tabular}{r}
  {\bf hep-ph/0010079}\\
  CERN-TH/2000-301 \\
\end{tabular}}

\def\beq{\begin{equation}}
\def\eeq{\end{equation}}

\def\bmat{\begin{array}}
\def\emat{\end{array}}
\def\barr{\begin{eqnarray}}
\def\earr{\end{eqnarray}}

\def\l{\left}
\def\r{\right}

\def\lsim{\raise0.3ex\hbox{$\;<$\kern-0.75em\raise-1.1ex\hbox{$\sim\;$}}}
\def\gsim{\raise0.3ex\hbox{$\;>$\kern-0.75em\raise-1.1ex\hbox{$\sim\;$}}}
\def\mf{{\cal M}_F}

\def\irad{{\cal I}_\tau}
\def\dmsq{\Delta m^2}
\def\np#1#2#3{           {\it Nucl. Phys. }{\bf #1} (#2) #3}  
\def\pl#1#2#3{           {\it Phys. Lett. }{\bf #1} (#2) #3}  
\def\pr#1#2#3{           {\it Phys. Rev. }{\bf #1} (#2) #3} 
\def\prl#1#2#3{          {\it Phys. Rev. Lett. }{\bf #1} (#2) #3}

\begin{document}

\title{No-go for exactly degenerate neutrinos at high scale ?}

\author
{Amol S. Dighe$^1$\footnote{amol.dighe@cern.ch} and 
Anjan S. Joshipura$^{1,2}$\footnote{anjan.joshipura@cern.ch} \\[.5cm]}
\address
{$^1$Theory Division, CERN, CH-1211 Geneva 23, Switzerland.\\
$^2$Theoretical Physics Group, Physical Research Laboratory, \\
Navarangpura, Ahmedabad, 380 009, India.}
\maketitle

\begin{abstract}
We show in a model independent manner that, if 
the magnitudes of Majorana masses of neutrinos
are exactly equal at some high scale,
the radiative corrections {\it cannot} reproduce the observed
masses and mixing spectrum at the low scale, irrespective of the
Majorana phases or the mixing angles at the high scale. 

\end{abstract}

\bigskip

The data from the solar and atmospheric neutrino experiments
can be explained through the mixing of three active neutrinos
with nonzero masses. The atmospheric neutrino solution needs
$\nu_\mu-\nu_\tau$ mixing with the corresponding
mass squared difference of 
$\dmsq_{atm} \approx 10^{-3} - 10^{-2}$ eV$^2$,
and a large mixing angle, $\sin^2 2\theta_{atm} > 0.8$
\cite{atm}. The solar neutrino solution needs the 
mixing of $\nu_e$ with a combination of $\nu_\mu$ and 
$\nu_\tau$ with a corresponding mass squared difference 
$\dmsq_\odot \lsim 10^{-4}$ eV$^2$, the mixing angle
may be small or large, depending on the particular
solution \cite{solar}.

Since $\dmsq_{atm} >> \dmsq_\odot$,
there are three distinct patterns for the neutrino masses
from the point of view of mass hierarchy.
We define the mass eigenstate $\nu_3$ as the one such that 
$|\dmsq_{31}| \approx |\dmsq_{32}| \approx 
\dmsq_{atm}$, and the states $\nu_1$ and $\nu_2$ as the
ones separated by $|\dmsq_{21}| \approx \dmsq_{\odot}$,
with $\dmsq_{31}$ and  $\dmsq_{21}$ having 
the same sign (here $\dmsq_{ij} \equiv |m_i|^2 - |m_j|^2$). 
With this convention, 
the possible patterns for the neutrino masses are  
({\it i}) completely hierarchical: $|m_1| \ll |m_2| \ll |m_3|$,
({\it ii}) partially degenerate: $|m_1| \sim |m_2| \gg |m_3|$,
also called as {\it inverted hierarchy}  
or ({\it iii}) completely degenerate: $|m_1| \sim |m_2| \sim |m_3|$.
In contrast to the first two cases, 
the common mass in the third
case can be near the direct limit on the electron neutrino mass as
obtained from the Kurie plot \cite{tritium}.
Such a mass can have its own signature
in the neutrinoless double beta decay \cite{doublebeta}.
In order for the neutrinos to contribute 
even a small fraction of the dark matter \cite{darkmatter} 
of the universe, degenerate neutrino masses are essential.

The pattern of mixing in the lepton sector
(which definitely has at least one large mixing angle,
as suggested by the atmospheric neutrino data) 
is very different from that in the
quark sector (where all the mixing angles are small).  
The presence of large mixing angles 
is another indication that the
neutrino masses may be partially or completely
degenerate (cases {\it ii} and {\it iii} above).
The origin of this degeneracy, as well as its breaking 
(which leads to the mass splittings among neutrinos) 
needs to be understood theoretically. Here, we
concentrate on the latter. 

The $SU(2)\times U(1)$ interactions 
in the standard model (and its supersymmetric
generalisations) break the symmetry among generations
explicitly through different Yukawa couplings.
The resulting radiative corrections
modify the neutrino mass matrices
while evolving from a high scale
to a low scale. The effects of radiative evolutions
on the masses and mixings have been studied in
\cite{general,specific,rad1}. 
The most economical  possibility would be to 
have {\it exactly} degenerate neutrinos at the high scale
(the seesaw scale, or the GUT scale, for example), and
let the radiative corrections lead to the required mass 
differences and mixing angles.
The {\it exact} degeneracy of neutrino masses at the high scale
may be obtained by a symmetry \cite{symm-ex}, 
which is broken later in the charged lepton sector.

The models with {\it exact} degeneracy at the high scale are
constrained by the masses and mixings observed at the low scale
(at the experiments). In the context of specific models,  
some constraints have been found.
Refs. \cite{lola,espinosa} have examined the consequences of the
bimaximal mixing pattern at the high scale 
which was motivated to suppress
the effect of large degenerate masses of neutrinos in 
the neutrinoless
double beta decay. Taking the model of Georgi and Glashow 
\cite{gg} as an example,
they show that the 
bimaximal pattern is unstable under radiative corrections. 
Moreover, one obtains the inverted hierarchy ($m_1>m_2> m_3$) 
for the renormalized masses in case of the MSSM, 
which rules out the MSW solution for the solar neutrinos. 
Ref. \cite{rathin} assumes $U_{e3}=0$ at the high scale, and
claims that the mixing pattern is unstable under radiative
corrections irrespective of the initial values of the angles.

The above arguments are not complete in ruling out the case of 
degenerate neutrinos for the following reasons. 
In the standard model, one does indeed obtain the {\it normal} 
mass hierarchy ($m_1 < m_2 < m_3$) for the
renormalized neutrino masses\footnote{
Ref. \cite{rathin} claim inverted hierarchy in case of the SM
rather than the MSSM, but this can be traced to their having 
the wrong sign for the 
radiative correction parameter $\epsilon$ to be defined later.}. 
Even in the case of the inverted hierarchy
(that one gets with MSSM), matter can play an important role in
the solution to the solar neutrino problem, 
and this possibility
is not experimentally ruled out \cite{darkside}.
Moreover, one need not
insist on the bimaximal mixing pattern if the common degenerate 
mass is not much larger than the experimental limit coming from 
the neutrinoless double beta decay. 
Finally, while the arguments of \cite{rathin} hinge on the
assumption of $U_{e3}=0$, nonzero values of $U_{e3}$ are allowed by
the experiments\cite{chooz}.

In this paper, we reinvestigate the viability of the 
{\it exact} degenerate spectrum in a model independent way. 
We show that (I) in any model with the magnitudes of the neutrino 
masses exactly equal at some high scale, the mixing
at the high scale can always be defined in  such a way that 
its structure is preserved in the process of evolution,
so that the issue of the stability of mixing angles 
does not arise. 
(II) However, irrespective of the structure of the original 
mass matrix, 
the degenerate spectrum at a high scale {\it cannot} lead to 
the observed masses and mixings at the low scale, as long as 
the charged lepton masses are generated by only one Higgs particle.
The results are valid irrespective of the model
(in particular, they hold for the SM and the MSSM),
the Majorana phases of the degenerate neutrinos
or their mixing angles at the high scale.

The most general neutrino mass matrix $M_{\nu}$ in the Majorana basis is a
symmetric
complex matrix and it can always be diagonalized by a unitary matrix
$U_0$ in the standard way:
\beq
U_0^T\;M_{\nu}\;U_0=D~~, 
\eeq
where $D$ refers to a diagonal matrix with real and positive masses. 
In particular, $D$ is proportional to the identity matrix for the 
exactly degenerate spectrum. 
It then follows that for degenerate neutrinos, 
the Majorana mass matrix  $\mf(X)$
at the high scale $X$ in the flavour basis 
(in which the charged lepton mass matrix is diagonal)
can always be written as
\beq
\mf(X) =  m U^\ast_0 U^\dagger_0 \equiv m~ V~~,
\label{u-udag}
\eeq  
where $m$ is a positive real number denoting the common mass
of the degenerate system and $V$ is a symmetric unitary matrix.
The freedom in defining the phases of the charged leptons
can be used to replace $V \to KVK$ without loss of generality,
where $K$ is a diagonal phase matrix.
We use this freedom to make the third row and the third column
of $V$ (and hence of $\mf(X)$) real and positive. 
The matrix $V$ can be
parametrised in terms of only two angles and a phase \cite{bronco}.

The radiative corrections modify $\mf$ as one evolves down  to
the low scale $x$. With only one Higgs giving masses to the 
charged leptons, the Yukawa couplings of the charged leptons
are hierarchical ($h_e : h_\mu : h_\tau = m_e : m_\mu : m_\tau
\approx 3 \cdot 10^{-4} : 6 \cdot 10^{-2} : 1$). 
In the limit of neglecting the electron and muon Yukawa couplings, 
the radiative modification of the mass matrix is given by
\beq
\mf(x) = \irad \mf(X) \irad ~~,
\label{i-mf-i}
\eeq
where $\irad \equiv Diag(1,1,\sqrt{I_\tau})$.
We define the radiative correction parameter $\epsilon$ through
$\sqrt{I_\tau}\equiv 1+\epsilon$.
The value of $\epsilon$ is determined by the model.
In particular, for the SM and the MSSM, $\epsilon$ can be
written in the form 
\beq
\epsilon\approx {\cal C} {h_\tau^2\over (4 \pi)^2}ln {X\over x}
\label{epsilon}~~, 
\eeq
where $h_\tau \equiv m_\tau/v$ is the tau Yukawa coupling 
in the standard model. The value of the constant ${\cal C}$ 
is $(1/2)$ and $(-1/\cos^2\beta)$ in the case of SM and MSSM 
respectively.
Note that the sign of $\epsilon$ is opposite in these two cases,
and hence the mass shifts in these two cases are in opposite directions.

Using (\ref{u-udag}), (\ref{i-mf-i}) and 
the unitarity of $V$, we get
\beq
\mf(x)\mf(x)^{\dagger}=m^2 I+  m^2 (I_{\tau}-1)
\l(\bmat{ccc}
V_{13}^2&V_{13}V_{23}&V_{13}V_{33}\sqrt{I_\tau}\\
V_{13}V_{23}&V_{23}^2&V_{23}V_{33}\sqrt{I_\tau}\\
V_{13}V_{33}\sqrt{I_\tau}&V_{23}V_{33}\sqrt{I_\tau}&1+V_{33}^2I_{\tau}\\
\emat\r)~~,
\label{commute}
\eeq
where $I$ is an identity matrix. All the entries in the above matrix are
real and positive due to
the phase convention we have chosen
(the third column of $V$ has real and positive elements). 
The matrix
appearing in the second term has one zero eigenvalue, 
indicating that one of
the originally degenerate eigenvalues remains 
unchanged in magnitude after the radiative corrections.
The other two masses are affected by the radiative corrections 
so that the eigenvalues of (\ref{commute}) are given to
leading order in $\epsilon$ by
\barr
|m_a|^2&=&m^2 \nonumber \\
|m_b|^2&=&m^2 (1 +4\epsilon \sin^2\tilde{\Psi}+O(\epsilon^2)) \nonumber \\
|m_c|^2&=&m^2 (1 +4 \epsilon \cos^2\tilde{\Psi}+O(\epsilon^2))~~,
\label{eval}
\earr
where the angle $\tilde{\Psi}$ is defined such that 
\beq
V_{33} = -\cos 2 \tilde{\Psi}~~.
\label{v33}
\eeq
The corresponding eigenvectors are given by the 
columns of the following 
matrix:
\beq
U_x  =   R_{12}(\Omega) R_{23}(\Psi) 
= \l( \bmat{ccc}
c_\Omega & s_\Omega c_{\Psi} & s_\Omega s_{\Psi} \\
-s_\Omega & c_\Omega c_{\Psi} & c_\Omega s_{\Psi} \\
0 & -s_{\Psi} & c_{\Psi}
\emat \r)~~, 
\label{u-expl}
\eeq
where $c_{\theta}\equiv\cos\theta,s_{\theta}\equiv\sin\theta$, and
the matrices $R_{ij}$ are the rotation matrices in the
corresponding planes. The angles $\Omega$ and $\Psi$ are
given by
\beq
\tan\Omega={V_{13}\over V_{23}}~~~~~; ~~~~~~~~~
\tan\Psi= (1 + 2 \epsilon) \tan\tilde{\Psi}+O(\epsilon^2) 
\label{angles} 
\eeq

The matrix $U_x$ (\ref{u-expl}) is the mixing matrix
at the low scale $x$. The corresponding matrix $U_X$ at the high scale is
defined only up to $U_X \to U_XOK$
due to the exact degeneracy in masses (here $O$ is an orthogonal matrix,
and $K$ is a diagonal phase matrix), however 
a ``natural'' choice of $U_X$ can be made.
Indeed, following the arguments in
\cite{bronco}, one can always write $V$ in the form
\barr
V & = &   
\l( \bmat{ccc} 
c_{\Omega'} & s_{\Omega'} & 0 \\
-s_{\Omega'} & c_{\Omega'} & 0 \\
0 & 0 & 1
\emat \r)
\l( \bmat{ccc}
e^{2 i\alpha} & 0 & 0 \\
0 & c_{2\Psi'} & -s_{2\Psi'} \\
0 & -s_{2\Psi'} & -c_{2\Psi'}
\emat \r)
\l( \bmat{ccc}
c_{\Omega'} & -s_{\Omega'} & 0 \\
s_{\Omega'} & c_{\Omega'} & 0 \\
0 & 0 & 1
\emat \r)~~, \\
 & & \nonumber \\
 & = &   R_{12}(\Omega') Diag(e^{i \alpha},1,1) 
R_{23}(\Psi') Diag(1,1,-1) 
R_{23}^T(\Psi') Diag(e^{i \alpha},1,1) 
R_{12}^T(\Omega')~~,
\label{mf-general}
\earr
which conforms to our phase convention of a real third row
and third column. 
The matrix $U_X$ that diagonalizes this $V$ 
(and hence, $\mf(X)$) can be chosen to be
\beq
U_X = R_{12}(\Omega') R_{23}(\Psi') = 
\l( \bmat{ccc}
c_{\Omega'} & s_{\Omega'} c_{\Psi'} & s_{\Omega'} s_{\Psi'} \\
-s_{\Omega'} & c_{\Omega'} c_{\Psi'} & c_{\Omega'} s_{\Psi'} \\
0 & -s_{\Psi'} & c_{\Psi'}
\emat \r)~~.
\label{u-X}
\eeq
This mixing matrix $U_X$ at the high scale 
has the same form as the matrix $U_x$ at the
low scale (\ref{u-expl}) resulting after the radiative
corrections. 
Expanding the right hand side of
(\ref{mf-general}), we get  $V_{33} = -\cos~ 2 \Psi'$, 
so that using (\ref{v33}), 
we can identify $\tilde\Psi = \Psi'$.
This implies that $\tan \Psi = \tan \Psi' [1 + O(\epsilon)]$.
Also,
$\tan \Omega' = \frac{V_{13}}{V_{23}}$,
so that from (\ref{angles}), we get $\Omega = \Omega'$.
Thus, the radiative corrections 
leave $\Omega'$ unchanged\footnote{
This general result has been proved in \cite{rad2}
in the case of the mixing matrix of the form
$U = R_{12} \cdot R_{23}$. Here we have shown
that $U$ can always be brought in this form in the present context.}, 
and modify the  angle
$\Psi'$ only at $O(\epsilon)$.
The  specific choice  for $U_X$ made in (\ref{u-X}) 
is thus a ``natural'' choice, such that the mixing is
not perturbed in the process of evolution. 

The specific structure of the mixing matrix $U_x$ (eq. (\ref{u-expl}))
is imposed in a model independent way just by the requirement
that the neutrinos be exactly degenerate at some high scale $X$.
This structure does not allow
large angle solutions to the solar neutrino problem without 
conflicting either with the atmospheric neutrino data \cite{atm}
or the CHOOZ constraints \cite{chooz}. 
The small angle solution is still allowed  (see \cite{rad2} 
for the detailed allowed region in angles $\Omega,\Psi$ in this case).
However, even in that case it is not possible to obtain the required 
mass squared differences, as we show below.

Note that the rows of $U_x$ in (\ref{u-expl}) 
are labelled by the flavour indices $e,\mu,\tau$,
while its columns are labelled by the mass eigenstates $a,b,c$. 
Reordering of the mass eigenvalues amounts to the 
interchange of columns of $U_x$.
We have denoted by $a$ the eigenvalue which remains unchanged and 
the corresponding eigenvector at low scale 
(which does not have any $\tau$ flavour component)
is given by the first column
of $U_x$. The other columns correspond to the eigenvalues
$|m_b|^2$ and $|m_c|^2$ respectively.
The neutrino mass squared differences are given from 
eq.(\ref{eval}) by
\beq
\dmsq_{cb}\approx 4 \epsilon  m^2 \cos 2\tilde{\Psi}~~,~~
\dmsq_{ca}\approx 4 \epsilon m^2 \cos^2 \tilde{\Psi}~~,~~
\dmsq_{ba}\approx 4 \epsilon m^2 \sin^2 \tilde{\Psi}~~. 
\label{deltas}
\eeq
It is seen from (\ref{deltas}) that two hierarchical
mass squared differences are possible if
({\em i}) $\tilde{\Psi}\sim \pi/4$, 
({\em ii}) $\tilde{\Psi}\sim \pi/2$, 
({\em iii}) $\tilde{\Psi}\sim 0$. 
These three cases correspond to 
the identification of the mass eigenstate
$\nu_3$ with $a,b,c$ respectively.
We shall show below that all the three identifications lead to 
phenomenological problems.
Note that since the arguments below depend only on the
ratios of the mass squared differences, they are
independent of the magnitude or sign of $\epsilon$,
and hence are valid for all the models.

In case ({\em i}), the first column of $U_x$ should be 
identified with the third column of the  
leptonic mixing matrix $U$, leading to the prediction
$$|U_{e3}|^2 + |U_{\mu 3}|^2 = 1~~.$$
This is clearly in contradiction with the required values for the
atmospheric neutrino mixing and the bound on $|U_{e3}|$ from CHOOZ
\cite{chooz}.

In case ({\em ii}), one has 
\barr
\frac{\Delta m^2_{\odot}}{\Delta m^2_{atm}} &\approx&
 \cot^2 \tilde{\Psi}~~, 
\label{ds/da} \\
\sin^2 2 \theta_{atm}&=&4 c_\Omega^2 c_{\Psi}^2 (1-c_\Omega^2 c_{\Psi}^2)
~~.
\label{theta-atm}
\earr
Since $\dmsq_\odot / \dmsq_{atm} < 0.1$, from (\ref{ds/da}) we have 
$\cos^2 \Psi \approx \cos^2 \tilde \Psi < \cot^2 \tilde \Psi < 0.1.$
Then eq.~(\ref{theta-atm}) gives $\sin^2 2 \theta_{atm} < 0.4$,
which is clearly inconsistent with the atmospheric neutrino data. 

In case ({\em iii}), 
\barr
\frac{\Delta m^2_{\odot}}{\Delta m^2_{atm}} &\approx&
\tan^2 \tilde{\Psi} ~~,
\label{ds/da3} \\
\sin^2 2 \theta_{atm}&=&4 c_\Omega^2 s_{\Psi}^2 (1-c_\Omega^2 s_{\Psi}^2)
~~.
\label{theta-atm3}
\earr
Since $\dmsq_\odot / \dmsq_{atm} < 0.1$, from (\ref{ds/da3}) we have 
$\sin^2 \Psi \approx \sin^2 \tilde \Psi < \tan^2 \tilde \Psi < 0.1.$
Then eq.~(\ref{theta-atm3}) gives $\sin^2 2 \theta_{atm} < 0.4$,
which is clearly inconsistent with the atmospheric neutrino data,
similar to the case ({\em ii}) above.

The above reasoning holds both in case of the small angle as well
as the large angle solution to the solar neutrino problem. 
As already remarked,
the latter can be independently ruled out from the structure of the 
mixing matrix $U_x$.
Thus, it is not possible to generate the observed pattern of
neutrino masses and mixings solely through radiative corrections,
starting from exactly degenerate Majorana neutrinos at some
high scale $X$. This conclusion is valid irrespective of the 
Majorana phases or the mixing angles at the high scale, or
the value of the high scale itself as long as there is no new
physics between the high and the low scale.

When the masses at the high scale  are not exactly equal,
but the mass splittings are small  compared to the magnitude of
radiative corrections ($\dmsq_{ij}(X)  \ll \epsilon |m_i|^2$),
the radiative corrections determine the mixing matrix at the low
scale, and hence the mixing matrix is $U_x$ 
as given by  eq.(\ref{u-expl}). 
The mass  squared differences at
the low scale are still dominantly given by eq.(\ref{deltas}).
(In the case where the original Majorana masses of the
neutrinos are $m_i = \pm m$, this is shown explicitly in \cite{haba}.) 
This implies that the required masses and mixings 
at the low scale still cannot be
reproduced even with the introduction of small mass splittings
($\dmsq_{ij}(X)  \ll \epsilon |m_i|^2$) at the high scale.
On the other hand, 
when $\dmsq_{ij}(X)  \gg \epsilon |m_i|^2$, the radiative
corrections clearly fail to have any significant impact on the
values of $\dmsq_{ij}$ and the mixing angles.
The net result is that at least one of the mass splittings
at the high scale ($\dmsq_{ij}(X)$) 
needs to be approximately equal to the one
observed at the low scale ($\dmsq_{ij}(x)$), 
and hence needs to be generated through some other mechanism.
The radiative corrections can help only in generating 
the other (most likely smaller) mass splitting.
Such masses and mixing patterns at the high scale have been
shown to lead to correct masses and mixings at the low scale
in several models \cite{models}.

The other possibility is that the degeneracy is completely broken 
at the high scale explicitly by a source other than the Yukawa
couplings. Specific models have been investigated in 
\cite{rathin,symm-br}. These may give rise to quasi-degenerate
neutrino masses at the high scale. One possible origin of such
splitting is the running of the degenerate right handed neutrino
masses from the Planck scale to the GUT or the scale of the right
handed neutrino masses\cite{espinosa}. The
radiative corrections
can then modify the mass splittings as well as the mixing angles.
The change in the mass splittings will be proportional to
the magnitude of the radiative corrections, but the mixing
angles can get modified drastically -- they can get magnified
or can be driven to zero, depending on the magnitude and
sign of the radiative corrections as compared to the
original mass splittings at the high scale 
\cite{general,specific,haba}. It may even be possible to obtain
large angles at the low scale irrespective of the 
mixing angles at the high scale \cite{rad1}.

Throughout this analysis, we have assumed that 
the charged lepton Yukawa couplings involve only one Higgs.
In the models in which two or more Higgs contribute 
to the charged lepton masses, the possibility of
exactly degenerate neutrinos at the high scale reproducing the
mass and mixing spectrum at the low scale is still open.

\vskip 1.0truecm
\noindent {\bf Acknowledgements}:\\
We thank J. Ellis for going through the manuscript. 
ASJ would like to thank the Theory Group at CERN for the kind hospitality
extended to him during the course of this work.

\end{document}